# Observation of the Josephson Plasma Mode for a Superfluid ³He Weak Link

A. Marchenkov, R. W. Simmonds, A. Loshak, J. C. Davis, and R. E. Packard

*Department of Physics, University of California, Berkeley, California 94720*

**The quantum phase difference across a superfluid Josephson-like weak link may exhibit periodic motion describable by a nonlinear equation identical to that of a rigid pendulum. We have directly observed this mode and find that the phase oscillation frequency decreases toward zero as the phase amplitude increases toward $\pi$. We also find that the low amplitude frequency is a direct measure of the weak link's critical current in quantitative agreement with theory.**

PACS numbers: 67.57.-z, 74.50.+r

One of the first published analysis of dynamical effects associated with a superfluid ³He Josephson-like weak link, predicted[1] unusual non-linear periodic motion of a coupled flexible membrane which is used to drive and detect mass current through the link. As we show below, a simple theory implies that the membrane-weak-link system will behave as a rigid pendulum with the quantum phase difference, $\phi$, playing the role of the pendulum's displacement angle. For instance, the frequency should depend on the oscillation amplitude, $\phi_{max}$, approaching zero when $\phi_{max} \to \pi$, while at low oscillation amplitudes the frequency of free-ringing motion is proportional to $\sqrt{I_c}$, where $I_c$ is the weak link's critical current. In this paper we report the first direct observations of these non-linear dynamical effects and the predicted $I_c$ dependence of the frequency. The latter result demonstrates that the low amplitude frequency of the free oscillations is a convenient measure of the critical current of the weak link. This is of interest to the operation of a superfluid dc-SQUID using two weak links[2].

Our weak link is a square array of 4225 apertures of 100 nm diameter, etched in a 50 nm thick SiN membrane. Since each aperture's diameter is on the order of the temperature dependent coherence length[3], ξ, they are expected to act as weak links. O. Avenel and E. Varoquaux originally reported[4,5] that the dynamics of an experimental apparatus containing a single aperture of nominal size 300 nm by 5000 nm in a 200 nm thick nickel film, could be successfully analyzed in terms of a sine-like current-phase relation for temperatures close to the superfluid transition temperature, $T_c$. Our experiments have shown[6] that the entire aperture array behaves coherently as a single weak link, analogous to a superconducting microbridge. The array exhibits a sine-like current phase relation[7] especially for temperatures $T>0.75T_c$.

The basic apparatus is similar to that used in our previous studies of the aperture array[6,7,8,9]. The apparatus includes a cylindrical inner cell bounded on the top by a flexible membrane $M_1$ with stiffness constant $k_1$ (force/displacement) and on the bottom by a stiffer membrane $M_2$ with stiffness constant $k_2$. The areas, $A$, of both membranes are the same. The weak link array is mounted near the center of $M_2$. Metallized plastic membrane $M_1$ is used both to induce the flow of the fluid through the aperture array, and to measure the amount of fluid motion. The fluid is pumped through the array by applying a voltage to the electrode, located next to $M_1$; a SQUID based position transducer[10] is used to measure the average membrane displacement from the equilibrium position and thus determine the flow.

In what follows we first ignore damping. The system's governing equations include the Josephson-like current phase relation:

$$I = I_c \sin f \qquad (1)$$

The Josephson-Anderson phase evolution equation:

$$\frac{df}{dt} = -\frac{2m_3 P}{\hbar r} \qquad (2)$$

the force balance equation for the membranes:

$$P = \frac{k_1 x_1}{A} = \frac{k_2 x_2}{A}, \qquad (3)$$

and the mass conservation equation:

$$I(t) = br\dot{x}_1(t)A \qquad (4)$$

In the equations above, $f$ is the quantum phase difference across the weak link, $r$ is the liquid density, $2m_3$ is the mass of a $^3$He Cooper pair, $P$ is the pressure difference across the weak link, $x_1$ and $x_2$ are the average displacements from the equilibrium positions of the top and the bottom membranes, respectively, and $b \equiv 1 + \dfrac{k_1}{k_2}$.

Differentiating Equation 2 and combining terms yields

$$\ddot{f} = -w_p^2 \sin f, \qquad (5a)$$

which implies that the equation governing the quantum phase difference $f$ is identical to the equation of a <u>rigid pendulum</u>. The small amplitude oscillation frequency is given by[11],

$$w_p^2 = \frac{2m_3 k_1}{b\hbar r^2 A^2} I_c \qquad (5b)$$

We have written Equation 5b in terms of the parameters of membrane $M_1$ since that is the membrane we monitor with the SQUID displacement transducer. Equation 5b displays the important relationship between critical current and frequency of the low amplitude pendulum-like oscillations: $w_p \propto \sqrt{I_c}$.

An impulsive force delivered to a rigid pendulum causes qualitatively different motion depending on the amplitude of the impulse[1,12]. 1) A small impulse excites simple harmonic motion at frequency $w_p$. 2) A larger impulse can excite anharmonic periodic motion. The amplitude dependent frequency decreases toward zero at a critical impulse that corresponds to the limiting oscillation amplitude $f_{max}=p$. For this special situation the restoring force approaches zero when the rigid pendulum moves toward the inverted position. 3) A still greater impulse causes the pendulum to rotate continuously in complete circles with circular frequency proportional to the impulse. Following a very large impulse, it is this latter type of motion which produces the previously reported[6] mass-current oscillations at the Josephson frequency (i.e. "Josephson oscillations"). By contrast, the small amplitude harmonic motion is analogous to the so-called plasma oscillation[13,14] that occurs due to the self-capacitance and intrinsic inductance of a superconducting Josephson weak link.

Since the membrane position, $x_1$, is proportional to $\dot{f}$, (Equations 2 and 3), these three qualitative features should be directly reflected in $x_1(t)$. In the present experiment we focus on smaller applied impulses than that required to send the system into the high frequency Josephson oscillation regime. We apply an impulse to the system by the application of a step voltage between the metallized membrane $M_1$ and the adjacent rigid electrode. The electrostatic attraction between the two electrodes creates a pressure step which excites the phase oscillation. Figure 1 shows a typical transient response of our experimental cell to an impulse. Here we have made the impulse sufficiently large so that the system has been pushed initially into the unbounded Josephson oscillation regime; i.e. the pendulum is rotating through angles greater than $2\pi$ and the average displacement, $<x_1> \propto <P> \neq 0$. As the energy dissipates, the system reaches a point (near $t \approx 1.05s$) where the $f$ oscillation amplitude is just below $\pi$ and the motion becomes bounded as pendulum oscillations commence, with $<x_1>=0$. For the first few cycles, the phase oscillation amplitude is large and the frequency[15] is small but, after a few cycles, as the amplitude decreases, the frequency increases, reaching an almost amplitude independent limit.

Since it is the equation for $f$ that represents pendulum motion, and $x_1 \propto \dot{f}$, the expected characteristic slowing of the "pendulum" near $f=p$, translates into a decrease in the slope of $x_1(t)$ near $<x_1>=0$. This slope change is clearly visible in Figure 1 in the first few cycles after the system falls into the pendulum mode. From large-amplitude phase oscillations such as shown in Figure 1, we have previously determined the current-phase relation, $I(f)$, of the array weak link[7]. In this paper, we analyze the dynamics of the transient response of the cell containing this weak link.

Figure 2 shows the pendulum frequency as a function of phase oscillation amplitude. The data shown are produced by averaging approximately 50 ring-down events at the same temperature. The expected decrease in frequency with increasing amplitude is clear. The smooth curve superimposed on the data is the analytic prediction[12] for a rigid pendulum without damping. By combining Equations 2 and 3, it is apparent that $f(t)$ can be found by integrating $x_1(t)$. There is a small error in $f$ due to neglecting damping. For more exact predictions we need to augment Equations 1-5 by including an appropriate damping term. Using numerical methods we find that the damping is sufficiently small that the simple interpretation that $x_1 \propto \dot{f}$ is negligibly changed. That is why the predicted amplitude dependent frequency for the undamped pendulum agrees so well with the data, as seen in Figure 2.

Equation 5b predicts that the low amplitude oscillation frequency, $w_p^2 \propto I_c$. In a previous paper[7] we have described how we determine the current-phase relation (by integration and differentiation of $x_1(t)$) and thereby find $I_c$. Using that method in this experiment we determine $I_c$ so that we can test Equation 5b. The correlation between $w_p$ and $I_c$ is shown in Figure 3. We see clearly the predicted proportionality between $w_p^2$ and $I_c$. In this figure we also plot Equation 5b, which **has no adjustable parameters**, and find it agrees very well with the data. The data shown are for the temperature regime where $I \propto \sin f$, i.e. $T \geq 0.75 T_c$. We have shown in previous work[7] that as the temperature decreases, the current-phase relation eventually becomes distorted from a simple sine function. Thus, when $T < 0.75 T_c$, as might be expected we find Equation 5b is no longer quantitatively correct[16].

Due to the excellent quantitative agreement between Equation 5b and the data in Figure 3, it appears that a direct measurement of the low amplitude oscillation frequency is a convenient method for determining $I_c$ in the Josephson regime. In the future it may be possible to make an analog of a dc-SQUID by placing aperture arrays in opposite arms of a superfluid quantum interferometer. Such a system should have an overall current-phase relation that is sine-like, with the maximum current being modulated by rotation flux through the arms of the device. Since the weak links described here are in the weakly damped regime, the traditional method of reading out a dc-SQUID (by applying a current larger than $I_c$ and detecting the chemical potential difference) may not be practical. However, from the results of the present experiment it would seem that the pendulum frequency of a superfluid dc-SQUID would also be modulated by the rotation flux[2] and thus provide a novel and convenient readout technique for the SQUID.

In conclusion, we have shown that a superfluid oscillator containing a Josephson-like weak link as the inertial element displays periodic motion similar to a weakly damped rigid pendulum. This is directly analogous to the Josephson plasma mode in superconducting weak links. The frequency is amplitude dependent in accord with simple theory. Furthermore, the quantitative connection between $w_p$ and $I_c$ permits a simple measure of the weak link critical current.

# Acknowledgements.

We wish to thank S. Pereversev who assembled the experimental cell, S. Backhaus and K. Shokhirev who performed analysis in an earlier version of this experiment and Prof. S. Vitale who has contributed many helpful suggestions. This research is sponsored by the NSF, NASA, NEDO and the ONR.

# Figures.

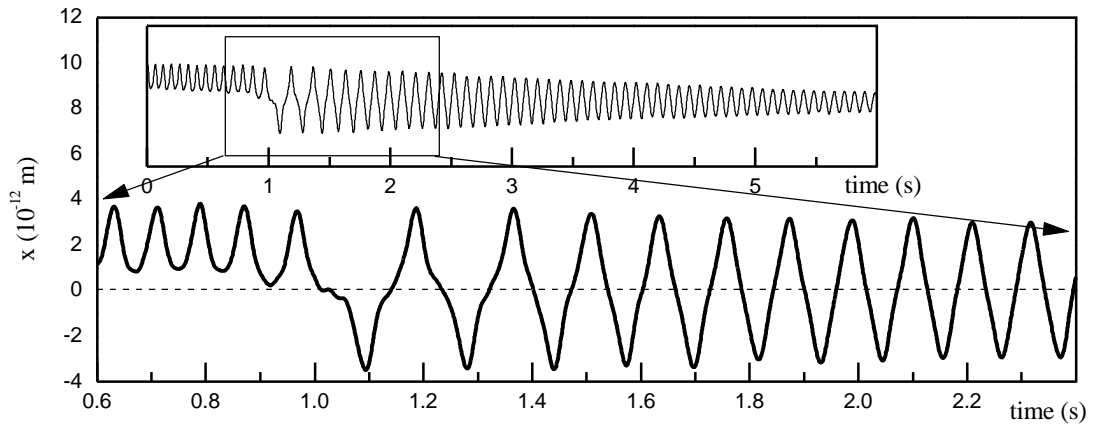

**Figure 1.** The position of membrane $M_1$ following the application of an impulse. During the first several oscillation cycles the system is in the Josephson oscillation regime and the average position (and hence average pressure difference across the weak link) is non-zero. At $t\sim 1.05$ sec the system becomes trapped in the bounded pendulum state. When the oscillation amplitude is large, the anharmonic motion near $<x_1>=0$ is characteristic of the rigid pendulum. The frequency can be seen to be increasing as the amplitude decays. This particular transient was observed at $T/T_c=0.85$ where the decay time is rather long and $I=I_c\sin f$.

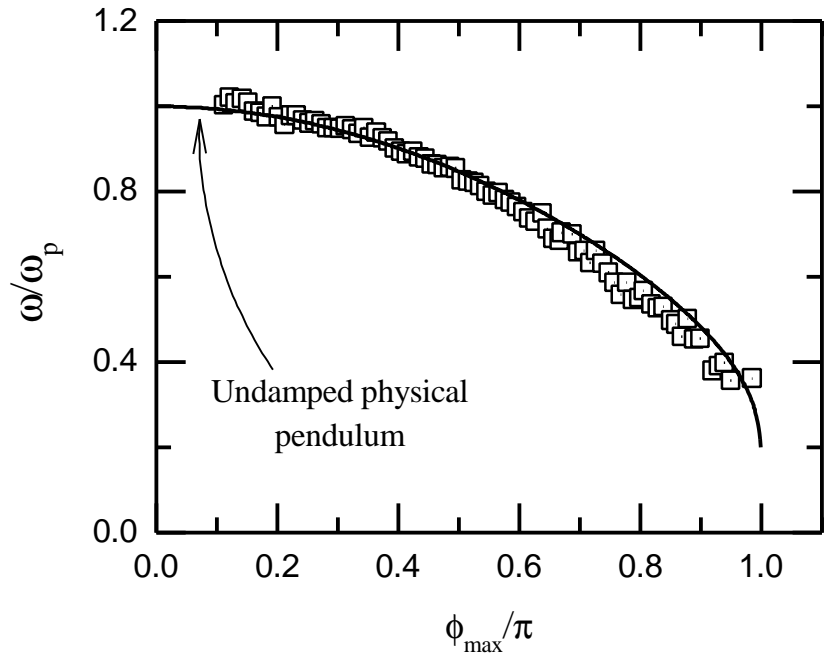

**Figure 2.** A plot of the oscillator frequency[15] as a function of phase angle. An ideal rigid pendulum (without damping) would follow the smooth curve drawn as a solid line[12]. The data are the average of approximately 50 transient ring-downs.

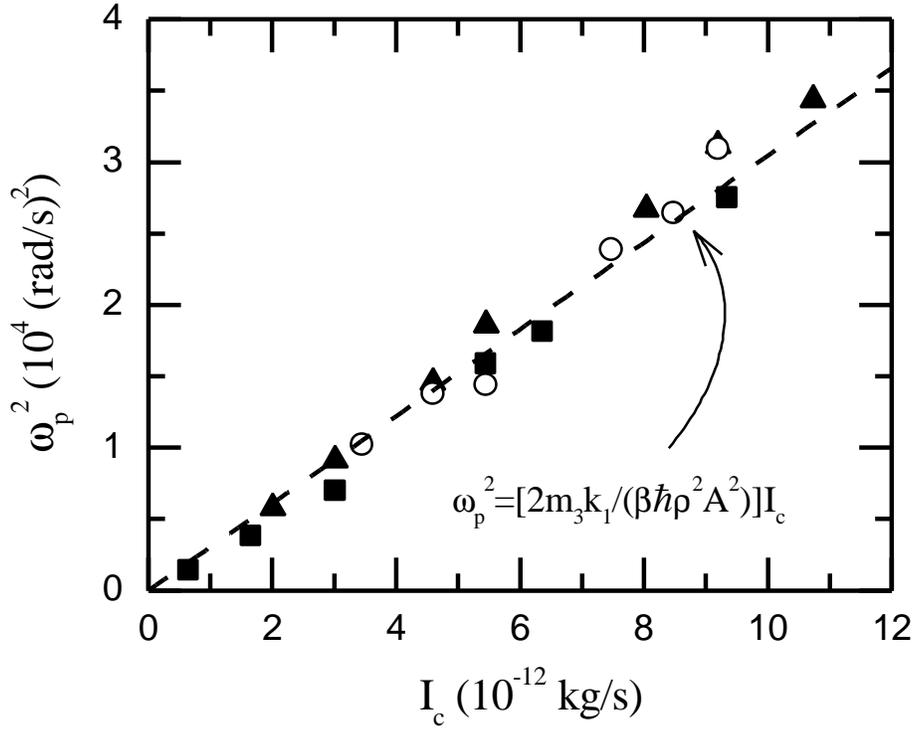

**Figure 3.** A plot of the dependence of the low amplitude (pendulum) frequency on the critical current. $I_c$ was determined by measuring the complete current phase relation in the temperature regime above $0.75T_c$, where $I(f)$ is sine-like. The straight line drawn is a plot of Equation 5b which has no adjustable parameters. At lower temperatures, when $I(f)$ is no longer a simple sine function, the frequencies drop below the line. Different symbols correspond to different cool-downs below $T_c$.